\documentclass{v23windows}
\usepackage{mathtools,amsmath,empheq,slashed}

\def\be{\begin{equation}}
\def\ee{\end{equation}}
\def\bea{\begin{eqnarray}}
\def\eea{\end{eqnarray}}

\newcommand{\fr}{\frac}
\newcommand{\eq}[1]{Eq.~(\ref{#1})}
\newcommand{\bib}[1]{Ref.~\cite{#1}}

\newcommand{\fig}[1]{Fig.~\ref{#1}}

\newcommand{\gev}{{\unskip\,\text{GeV}}}

\begin{document}
\vspace*{4cm}
\title{Broken scale invariant unparticle physics and its prospective effects on the MuonE experiment}

\author{\underline{Van Dung Le}$^{a,b}$, Duc Ninh Le$^c$, Duc Truyen Le$^d$, Van Cuong Le$^{a,b}$}


\address{
$^{a}$Department of Theoretical Physics, University of Science, Ho Chi Minh City 70000, Vietnam\\
$^{b}$Vietnam National University, Ho Chi Minh City 70000, Vietnam\\
$^{c}$Faculty of Fundamental Sciences, PHENIKAA University, Hanoi 12116, Vietnam\\
$^{d}$Department of Physics, National Tsing Hua University, Hsinchu, Taiwan 300044, R.O.C.}

\maketitle\abstracts{
We investigate the effects of broken scale invariant unparticle at the MUonE experiment. The choice of the broken model is because the original scale-invariant model is severely suppressed by constraints from cosmology and low-energy experiments. Broken scale invariant unparticle model is categorized into four types: pseudoscalar, scalar, axial-vector, and vector unparticle. Each uparticle type is characterized by three free parameters: coupling constant $\lambda$, scaling dimension $d$, and energy scale $\mu$ at which the scale-invariance is broken. After considering all of the available constraints on the model, we find that the MUonE experiment is sensitive to (axial-)vector unparticle with $1 < d < 1.4$ and $1\le \mu \le 12$ GeV.
}

\section{Broken scale invariant unparticles and constraints}
Unparticle model was proposed by Georgi in 2007 to discuss the physics of the scale invariant sector in the infrared region \cite{Georgi:2007ek}. 
 This unparticle stuff, which can be viewed as a set of $d$ invisible massless particles with $d$ being a non-integral number, 
 interacts with the Standard Model (SM) particles through effective interactions of the form
\begin{equation}
\frac{C_\mathcal{U}}{\Lambda_\mathcal{U}^{d}} \mathcal{O}_{SM} \mathcal{O}_{\mathcal{U}},
\label{L_Unparticle}
\end{equation}
where $\Lambda_\mathcal{U}$ is the energy scale at which the unparticle emerges. 
The parameter $d$ is the scaling dimension of the unparticle operator $\mathcal{O}_{\mathcal{U}}$. 
$C_\mathcal{U}$ is a coupling parameter. These parameters are all unknown, so one could re-express them in terms of two 
parameters $\lambda$ and $d$. The scaling dimension $d$ must be kept because it appears also in the unparticle propagator. 

In this work we are interested in unparticle effects at the MUonE experiment \cite{CarloniCalame:2015obs,Abbiendi:2677471}, where the elastic scattering $e\mu \to e\mu$ cross sections are measured, 
the dominant contribution comes from the interactions between an unparticle and the charged leptons (electron or muon). We assume here lepton universality and no flavor-number violation for simplicity. 
The relevant SM operator in \eq{L_Unparticle} is therefore $\mathcal{O}_\text{SM} = \overline{f} \Gamma f$ 
with $\Gamma=$ $1$, $\gamma_5$, $\gamma_\mu$, $\gamma_\mu \gamma_5$. 
We will consider these cases separately. We note that the unparticles can couple to 
other SM fields such as the quarks, the gauge bosons, and the Higgs boson. These effects are however much weaker, hence are here neglected.

To be specific, we consider the following four operators \cite{Le:2023ceg}
\begin{equation}
  \frac{\lambda_S}{M_Z^{d-1}}\overline{f} f \mathcal{O}_\mathcal{U},\quad 
\frac{i\lambda_P}{M_Z^{d-1}}\overline{f} \gamma_5 f \mathcal{O}_\mathcal{U},\quad
\frac{\lambda_V}{M_Z^{d-1}}\overline{f} \gamma_\mu f \mathcal{O}_\mathcal{U}^\mu,\quad 
\frac{\lambda_A}{M_Z^{d-1}}\overline{f} \gamma_\mu\gamma_5 f \mathcal{O}_\mathcal{U}^\mu,
\label{vertices_lambda}
\end{equation}
which are called scalar (S), pseudo-scalar (P), vector (V), and axial-vector (A) unparticles, respectively. 

\begin{figure}[h!]
  \centering
  \includegraphics[width=0.6\textwidth]{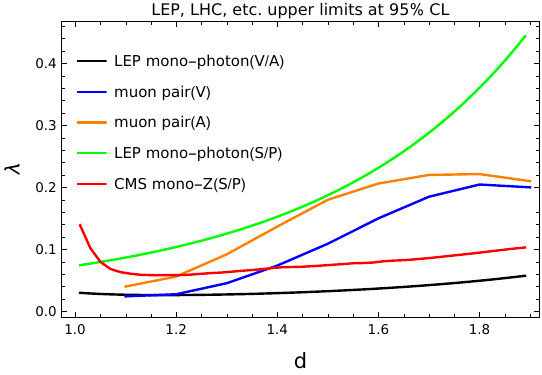}
  \caption{Upper limits at $95\%$ CL from LEP, CMS and other experiments data on (axial-)vector and (pseudo-)scalar unparticle parameters. The regions above the curves are excluded.}
  \label{bound_U}
\end{figure}

The original idea of Georgi suggests that the scale-invariant unparticles exist at the energy range $E \le \Lambda_\mathcal{U}$.   
It was however very soon realized that data from cosmology and low-energy experiments puts severe limits on the couplings between the unparticles and the SM sector, see e.g. \cite{Davoudiasl:2007jr,Liao:2007bx,Balantekin:2007eg,Barger:2008jt}, making it impossible to observe unparticle effects at present or near-future experiments. A simple way to evade these constraints was to introduce a scale breaking parameter $\mu$, which assumes that the contribution from the broken phase (i.e. energies less than $\mu$) is suppressed \cite{Fox:2007sy}. This slightly affects the unparticle propagators \cite{Fox:2007sy,Barger:2008jt}
\begin{eqnarray}
\Delta_F(k)&=&\frac{iZ_{d}}{(-k^2+\mu^2-i\epsilon)^{2-d}},\quad \text{(pseudo-)scalar unparticles},\label{prop-scalar-b}\\
\Delta^{\mu \nu}_F(p)&=&\frac{iZ_{d}}{(-k^2+\mu^2-i\epsilon)^{2-d}}\left(-g^{\mu\nu}+\frac{k^\mu k^\nu}{k^2}\right),
\quad \text{(axial-)vector unparticles},
\label{prop_vec_b}
\end{eqnarray}
where $k$ is the momentum of the unparticle,
\begin{align}
A_{d}=\frac{16\pi^{5/2}}{(2\pi)^{2d}}\frac{\Gamma(d+1/2)}{\Gamma(d-1)\Gamma(2d)}, \quad Z_{d} = \fr{A_d}{2\sin(d \pi)}.
\end{align}
The full scale invariance case as originally proposed by Georgi is recovered in the limit $\mu \to 0$. 
The Big Bang Nucleosynthesis (BBN) and SN 1987A constraints can be evaded by simply choosing a sufficiently large $\mu$, namely 
$\mu \ge 1$~GeV \cite{Barger:2008jt}. 
After considering all current experimental constraints for unparticle, we came up with the bounds for the interested parameters in Fig. \ref{bound_U}, which is taken from \bib{Le:2023ceg}.
\section{MUonE experiment and unparticle effects}
The aim of the MUonE experiment is to provide an independent and precise determination of the leading hadronic contribution to the muon anomalous magnetic moment using the following equation \cite{CarloniCalame:2015obs}
\begin{eqnarray}
  a_\mu^{\rm had} = \frac{\alpha(0)}{\pi} \int_0^1 dx(1-x)  \Delta \alpha_{\rm had}[t(x)],\quad 
  t(x) = -\fr{m_\mu^2 x^2}{1-x} < 0,
  \label{eqn:a_mu}
\end{eqnarray}
where $\Delta \alpha_{\rm had}(t)$ is extracted from the measurement of the running $\alpha(t)$ in the space-like region as
\begin{eqnarray}
\alpha(t) = \frac{\alpha(0)}{1-\Delta\alpha(t)} , \quad  
\Delta \alpha(t) = \Delta \alpha_{\rm lep}(t) + \Delta \alpha_{\rm had}(t),
\end{eqnarray}
with $\Delta \alpha_{\rm lep}(t)$ being the SM value.

The MUonE experiment measures precisely the following differential cross section, using the SM parametrization
\begin{align}
\frac{d \sigma_\text{SM}}{dT}=\frac{\pi \alpha^2(t)}{(E_\mu^2-m_\mu^2)m_e^2 T^2}[2E_\mu m_e(E_\mu-T)-T(m_e^2+m_\mu^2-m_e T)],\label{SMcrosssection}
\end{align}
where 
\begin{align}
t = -2m_e T = (p_\mu - p'_\mu)^2, \quad T = E'_e - m_e \ge 0 ,\label{def_t_T}
\end{align}
with $p_\mu$ and $p'_\mu$ being the momentum of the initial-state and final-state muons, respectively. 
$E'_e$ is the energy of the final-state electron in the laboratory (Lab) frame. The variable $T$ is essentially 
$E'_e$ in practice.
The energy of the incoming muon in the laboratory frame is $E_\mu = 150\gev$. The center-of-mass energy is 
$\sqrt{s} = \sqrt{2E_\mu m_e + m_\mu^2 + m_e^2} \approx 0.4\gev$. Because of this low center-of-mass energy, the 
contribution from the $Z$ boson is negligible and has been removed from \eq{SMcrosssection}. 
The design of the experiment is to measure the $d\sigma/dT$ distribution at the level of $10$~ppm systematic uncertainty \cite{Abbiendi:2677471}. The purpose of this work is to check whether unparticles effects can be detected at this level of accuracy.  
We then need to calculate the unparticle contributions to the differential distribution $d\sigma/dT$. 
Analytical results for four unparticle cases can be found in \bib{Le:2023ceg}. We summarize only the important numerical results in the following. These results are taken from \bib{Le:2023ceg}. 

\begin{figure}[h!]
  \centering
  \begin{tabular}{cc}
  \includegraphics[width=0.49\textwidth]{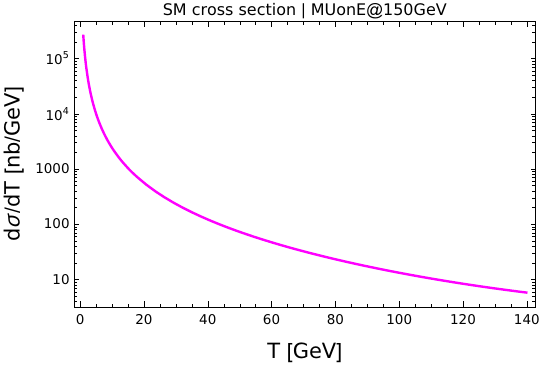} 
  \includegraphics[width=0.49\textwidth]{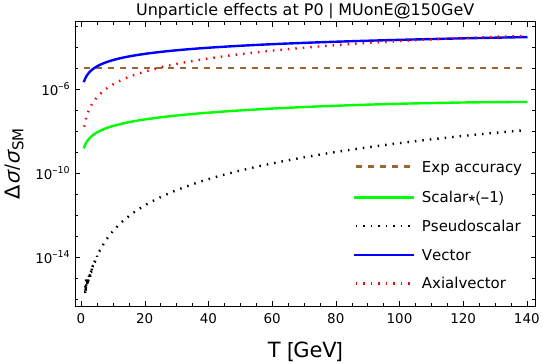}\\ 
  \end{tabular}
  \caption{Left: Differential cross-section of the SM. Right: Various unparticle effects calculated at the parameter point P0 relative to the SM values. The MUonE systematic accuracy level of $10$ ppm is indicated by the dashed brown line.}
  \label{fig:SM_U_P0}
\end{figure}

To demonstrate the effects of unparticles on the differential cross-section, we choose the following benchmark point 
\begin{align}
\text{P0}:\quad d=1.1,\quad \lambda_i=0.02, \quad \mu=1\gev, 
\end{align}
where $i=S,P,V,A$. This point $\text{P0}$ satisfies all the constraints presented in \fig{bound_U}. 
The results are shown in \fig{fig:SM_U_P0}. 
We see that the effects of the (axial-)vector unparticles can be visible at the 
the MUonE experiment, while those of the (pseudo-)scalar unparticles seem too weak to be detected.  

\begin{figure}[h!]
  \centering
  \includegraphics[width=0.49\textwidth]{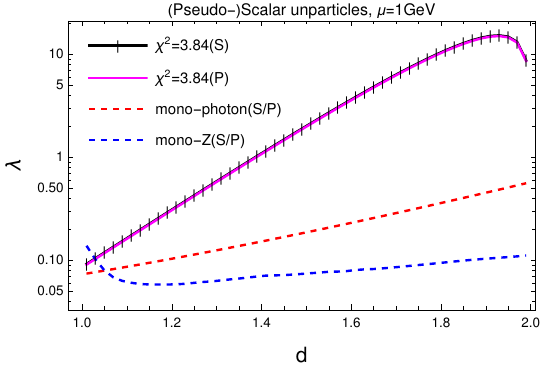}
  \includegraphics[width=0.49\textwidth]{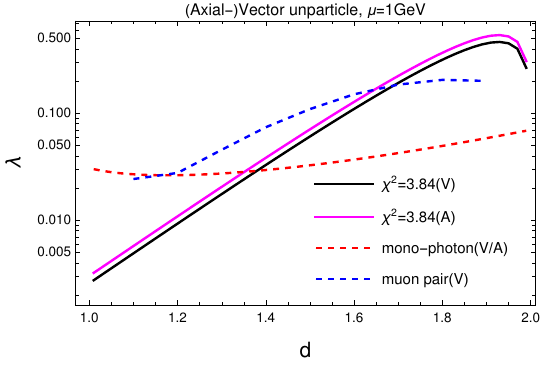}\\
  \includegraphics[width=0.49\textwidth]{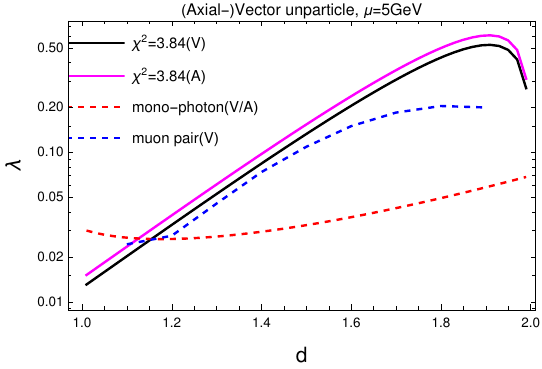}
  \includegraphics[width=0.49\textwidth]{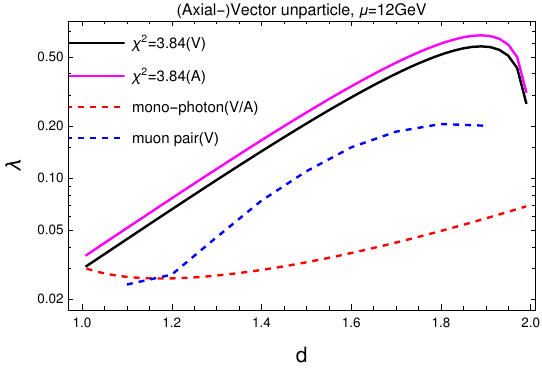}\\
  \caption{Sensitivity curves of the MUonE experiment on the unparticle-SM coupling $\lambda$ 
  for the cases of scalar (S), pseudo-scalar (P), vector (V), and axial-vector (A) unparticles. 
  The $95\%$ CL upper limits from the mono-photon, mono-$Z$, and muon-pair productions are also plotted. 
  The muon-pair bound for the pseudo-scalar case is not plotted as it is irrelevant.}
  \label{sens_curves}
\end{figure}

To have firmer conclusions, we plot the sensitivity curves in \fig{sens_curves} scanning over the parameter 
space of $(d$, $\lambda$, $\mu)$.
The $\chi^2$ function is given in \bib{Le:2023ceg}. 
From this we conlcude that the MUonE experiment is insensitive to the (pseudo-)scalar unparticles, but it is 
sensitive to the vector unparticle if $1 \le \mu \le 12$~GeV and $1 < d < 1.4$. The sensitivity to the 
axial-vector unparticle is similar, albeit a bit weaker. We also observe that the vector unparticles can 
affect significantly the best-fit value of $a_\mu^{\rm had}$.  
\section{Conclusions}
The MUonE experiment promises a novel approach to evaluating the hadronic contribution to the muon $(g-2)$. Such a precise experiment can help us to detect small new physics effects such as unparticles. From our analysis, unparticles with broken scale invariance are still possible. We found that MUonE is sensitive to the (axial-)vector unparticles with $1 < d < 1.4$ and $1 < \mu < 12 $GeV, while the effects of (pseudo-)scalar unparticles are too feeble to be detected.
\section*{Acknowledgement}
This research is funded by the  Vietnam National Foundation for Science and Technology Development (NAFOSTED) under grant number 103.01-2020.17.
\section*{References}
\bibliographystyle{unsrt}    

\end{document}